\documentclass[twocolumn,secnumarabic,nobalancelastpage,nofootinbibt]{revtex4}

\usepackage{graphics}      
\usepackage{graphicx}      
\usepackage{longtable}     
\usepackage{url}           
\usepackage{bm,color}            
\usepackage{amssymb, amsmath, enumerate, theorem, epsfig,subfigure,setspace}
\usepackage{amsfonts}
\begin{document}

\title{Discrete Polynomial Optimization with Coherent Networks of  Condensates and Complex Coupling Switching }
  \author{Nikita Stroev$^1$ and Natalia G. Berloff$^{1,2}$ }
\email[correspondence address: ]{N.G.Berloff@damtp.cam.ac.uk}
\affiliation{$^1$Skolkovo Institute of Science and Technology, Bolshoy Boulevard 30, bld.1,
Moscow, 121205, Russian Federation}
\affiliation{$^2$Department of Applied Mathematics and Theoretical Physics, University of Cambridge, Cambridge CB3 0WA, United Kingdom}


\begin{abstract}{Gain-dissipative platforms consisting of lasers, optical parametric oscillators and nonequilibrium condensates operating at the condensation/coherence threshold have been recently proposed as efficient analog simulators  of 2-local spin Hamiltonians with continuous or discrete  degrees of freedom.  We show that nonequilibrium condensates {\it above the threshold} arranged in an interacting network may realise $k$-local Hamiltonians with $k>2$ and lead to nontrivial phase configurations. The principle of the operation of such a system lays the ground for physics-inspired computing and the new efficient methods for finding  solutions   to  the higher order binary optimization problems. We show how to facilitate the search for the global solution by invoking complex couplings in the system and demonstrate the efficiency of the method on tensors with million entries. This approach offers a highly flexible new kind of computation based
on gain-dissipative simulators with complex coupling switching. }  
\end{abstract}

\maketitle

Recently much effort has been devoted to the development of various technological  platforms that act as  quantum or classical analog simulators aimed at solving certain classes of hard classical optimization problems \cite{utsunomiya2011mapping,marandi2014network,inagaki2016large,mcmahon2016fully,takeda2017boltzmann,nixon2013observing,dung2017variable}. It is expected that these kinds of platforms would help to efficiently solve many tasks  of significant computational complexity, ranging from modelling microscopic effects and processes like the behavior of electrons in complex materials \cite{buluta2011natural,georgescu2014quantum} and finding the ground state of spin glasses \cite{barahona1982computational}, to the applied combinatorial optimization problems  \cite{lawler1985traveling}. Large scale  computational problems of this type are hard for classical von Neumann architecture which suggests looking for fully analog or hybrid digital/analog/quantum  devices that can find a solution faster or find a better solution in a fixed time. 

Nonequilibrium condensates, optical parametric oscillators, lasers, and memristors have  been considered as annealing-inspired accelerators  and demonstrated successes in finding the ground state of  spin Hamiltonians with continuous or discrete variables \cite{marandi2014network, nixon2013observing,berloff2017realizing, memristors}. 
In particular, the Coherent Ising Machine  has been shown to significantly outperform classical simulated annealing in terms of both accuracy and computation time to efficiently solve Max-CUT problems \cite{marandi2014network}  and has shown better scalability than the quantum annealers \cite{hamerley}. Memristors with massively parallel operations performed in a dense crossbar array were shown to be able to solve NP-hard Max-CUT problems predicting  over four orders of magnitude advantage over digital and optical hardware \cite{memristors}. Integrated photonic circuits that use self-phase modulation in two microring resonators were shown to act  as an optical coherent Ising machine \cite{oic16,oic19}. The lattices of exciton-polariton condensates  were shown to efficiently simulate the XY Hamiltonian when operating at the condensation threshold \cite{berloff2017realizing,kalinin2018networks} 
In all these systems, discrete Ising 'spins' or continuous XY 'spins' are encoded in individual phase modes of the nonlinear  networks. An optimization problem of interest is mapped  into the connectivity matrix of the spin network with the task of finding its ground state, which can be related to finding the 'maximum occupancy' of the collective supermode of the underlying network, as a system specific gain mechanism is continuously increased to reach the coherence threshold  \cite{wang13,berloff2017realizing}.

The development of technological platforms that promise to offer a significant time or power consumption improvements in solving hard optimization problems goes hand in hand with annealing inspired optimization, when the physical principle of the device is used to formulate the classical algorithms \cite{aramon2019physics}. The recent examples of which include Simulated Bifurcation algorithms inspired by quantum adiabatic optimization using a nonlinear oscillator network \cite{toshibaBM}, destabilization of local minima based on degenerate parametric oscillator networks \cite{leleu19}, parallel tempering Monte Carlo \cite{rozada2019effects}, and the gain-dissipative algorithm based on operation of the polariton graph simulator at the condensation threshold \cite{kalinin2018simulating}.



The focus of all these technological and inspired implementations of the annealer-based optimization has been on QUBO, however, there is a large class of optimization problems ---  the higher order polynomial binary optimization (HOBO) -- that  are more naturally encoded by the $k$-local Hamiltonians \cite{he2013approximation,jiang2014approximation}. HOBO is concerned with optimizing a (high degree) multivariate polynomial function in  binary variables. Our basic model is to maximize or minimize  a $k$-th degree polynomial function $f({\bf x})$ where ${\bf x}=(x_1,..., x_i,..., x_N)$, $x_i\in\{-1,+1\}$. The examples of HOBO are ubiquitous from Hypergraph max-covering problem to Frobenius and "market split" problems \cite{jiang2014approximation}.  HOBO is a fundamental problem in integer programming and is also known as Fourier support graph problem. Any HOBO can be mapped into the QUBO \cite{de2016simple}, however, the  overhead in the number of nodes becomes prohibitive in an actual technological platform, so it is important to consider ways to solve HOBO directly. 
The purpose of this article is three-fold. First, we show that Ising machines based on nonequilibrium condensates can be used to address 4-local HOBO when operating {\it above} the threshold. 
Secondly, inspired by the operation of the networks of nonequilibrium condensates we  propose a new optimization algorithm for solving HOBO of arbitrary degree. Finally, we show that another physics-inspired method of turning on and off the complex coupling between the nonlinear condensates greatly enhances the search for global minimum.

{\it Polynomial optimization with coherent networks.} 
The  optimization problem studied in this paper  is 
\begin{equation}
\min_{{\bf x}\in\{-1,+1\}^N} -\sum_\Omega {\bf A}_{i_1,i_2,\cdot\cdot\cdot, i_k}^k x_{i_1} x_{i_2}\cdot\cdot\cdot x_{i_k},
\label{optimisation}
\end{equation}
where $\Omega=\{i_j: 1\le i_1 \le i_2 \le...\le  i_k \le N\}$ and ${\bf A}^k$ is the super-symmetric tensor of degree $k$. 
To formulate the gain-dissipative platform that reaches  the ground state of HOBO by  finding the 'maximum occupancy' collective supermode of the underlying network of nonequilibrium condensates we consider the mean-field equations that govern such a network based on  the Ginzburg-Landau equation \cite{wouters2007excitations, keeling2008spontaneous}. This is a universal  driven-dissipative equation that describes the behaviour of systems in the vicinity of a symmetry--breaking instability and has been used  to describe  lasers,  thermal convection, nematic liquid crystals, and various non-equilibrium condensates \cite{carusotto2013quantum, keeling2011exciton}. When  derived asymptotically from a generic laser model given by Maxwell-Bloch equations it has a saturable nonlinearity and  can be written  as    
\begin{equation}
{\rm i} \frac{\partial \psi}{\partial t} = - \nabla^2 \psi +  \tilde{U} |\psi|^2 \psi  +  i \biggl(\frac{P({\bf r},t)}{1 + b |\psi|^2}-\gamma_c\biggr) \psi,
\label{gpe}
\end{equation}
where $\psi({\bf r},t)$ is the wavefunction of the system,  $\tilde U$ is the strength of the delta-function interaction  potential, $\gamma_c$ is the rate of linear losses, $b$  parametrizes  the effective strength of nonlinear losses, $P({\bf r},t)$ describes the gain  mechanism that adds particles to the system. It was experimentally  demonstrated \cite{berloff2017realizing} that when pumped at the condensation threshold, freely expanding optically imprinted polariton condensates arranged in a lattice may achieve a steady state with condensate phases realising the minimum of the XY Hamiltonian. In this framework, the coupling strengths between condensates depend of the system parameters, pumping intensity and shape and on the lattice geometry \cite{lagoudakis2017polariton}. We shall assume that $P({\bf r},t)$ adds  particles in $N$ spatial locations centered at ${\bf r}_i$, $i=1,...,N$, so that $P({\bf r},t) =\sum_i f_i(t) p_i({\bf r})$, where $f_i$ is the time-dependent part of the pumping at ${\bf r}={\bf r}_i$ and  $p_i({\bf r})\equiv p(|{\bf r}-{\bf r}_i|)$ is a given spatially localised pumping profile, that creates the condensate with a wavefunction $\phi_i({\bf r})\equiv\phi(|{\bf r}-{\bf r}_i|)$ centred at ${\bf r}={\bf r}_i$  and normalized  so that $2\pi \int |\phi(r)|^2 r\, dr=1$.  In writing Eq.~(\ref{gpe}) we let $\hbar=1$ and $m=1/2$.  
If the distances between the neighbouring condensates are larger than the width of  $p(r)$, we employ the tight binding approximation and write the wavefunction of the system  as  a linear superposition of the wavefunctions of individual coherent centers 
$
\psi({\bf r},t) \approx \sum_{i=1}^N a_i(t) \phi_i({\bf r})
$,
where $a_i(t)$ is the time-dependent complex amplitude networks \cite{smerzi2003nonlinear, kalinin2018simulating}. We expand the first term
in the brackets of Eq.~(\ref{gpe}) in Taylor series, substitute the
expressions for $P$ and $\psi$, multiply by $\phi_j^*$ for $j=1,...,N$ and eliminate the spatial degrees of freedom by integrating in the entire space to obtain  $N$ equations of the form 
\begin{eqnarray}
\frac {d{\Psi }_{{i}}}{dt}&=&{\Psi }_{{i}}\left({{\gamma }_{{i}}-(\sigma_i+{\rm i}{U}){{\mid{{\Psi}_{{i}}}\mid}}^{{2}}}\right)+{\sum_{j,j\ne i} {{J}_{{ij}}{\Psi }_{{j}}}}\nonumber\\
&+&{\sum_{\langle j,k,l \rangle}{Q}_{{ijkl}}}{{\Psi }_{{j}}{\Psi }_{{k}}{\Psi }_{{l}}^*.}
\label{ai_alt}
\end{eqnarray}
In writing Eq.~(\ref{ai_alt}) we used the following notations:  $\gamma_i=f_i\int p|\phi|^2\, d{\bf r}-\gamma_c$, $\sigma_i=f_i\int p|\phi|^4\, d{\bf r}$, $U=\tilde{U}\int |\phi|^4\, d{\bf r}$, 
$
Q_{ijkl}=-b{{\sum _{m\in\{i,j,k,l\}}f_m\int{{p}_{{m}}{\phi }_{{k}}{\phi }_{{l}}^{{* }}{\phi }_{{i}}{\phi }_{{j}}^{{* }}}}} {d{\bf r}},
$
and $\langle i,j,k  \rangle$ denotes the combinations of $\{j,k,l\}$ that exclude $j=k=l=i$.
To the leading order  we kept the terms   $J_{ij}= Re[f_{i}{\int {{p}_{{i}}{\phi }_{{i}}{\phi }_{{j}}^{{* }}}} {d{\bf r}}+f_{j}{\int {{p}_{{i}}{\phi }_{{j}}{\phi }_{{i}}^{{* }}}} {d{\bf r}}]$ for $j\ne i,$   and 
 denoted $\Psi_i=a_i \exp({\rm i} t Re\int \phi^* \nabla^2 \phi\, d{\bf r})$. 
We rewrite Eq.~(\ref{ai_alt})  in terms of the number densities $\rho_i$ and phases $\theta_i$, $\Psi_i=\sqrt{\rho_i} \exp[{\rm i} \theta_i],$ 
$\frac{1}{2}\dot{\rho}_i(t)=(\gamma_i -\sigma_i \rho_i) \rho_i + \sum_{j, j\ne i} J_{ij} {\sqrt{\rho_i\rho_j}}\cos\theta_{ij} + \sum_{\langle j,k,l \rangle} Q_{ijkl} {\sqrt{\rho_i\rho_j\rho_k\rho_l}}\cos\theta_{ijkl},$ 
 %
 and
$\dot{\theta}_i(t)=-U\rho_i -\sum_{j,j\ne i} J_{ij} {\sqrt{\rho_j/\rho_i}} \sin\theta_{ij}-\sum_{\langle j,k,l \rangle} Q_{ijkl} {\sqrt{\rho_j\rho_k\rho_l/\rho_i}} \sin\theta_{ijkl},$ 
where $\theta_{ij}=\theta_{i}-\theta_{j}$ and $\theta_{ijkl}=\theta_{i}+\theta_{l}-\theta_{k}-\theta_{j}$.

The higher order terms affect the states even in the simplest configuration of two identical oscillators pumped with $\gamma_i=\gamma$ for which the  occupancy $\rho_{\rm 0}=\rho_1=\rho_2$ at the fixed point  reads
$\rho_{\rm 0}=[\gamma +  J\cos \Delta \theta + \tilde{ Q}\cos(2\Delta \theta)]/\sigma$  where $\Delta \theta=\theta_1-\theta_2$ and $\tilde{Q}=\rho_{\rm 0} Q.$  By  choosing the minimum pumping $\gamma$ to reach the required occupancy, we minimize the Hamiltonian $H_{\rm two}=-J\cos \Delta \theta - \tilde{Q}\cos(2\Delta \theta)$ while the equation on $\theta_i$  describes the gradient descent to the local minimum of $H_{\rm two}$ with the dynamics of the higher order Kuramoto oscillators. If $\tilde{Q}$ is negligible (close to condensation threshold), we have the minimization of the XY Hamiltonian, so $\Delta \theta=0$ or $\pi$ if $J>0$ or $J<0$ respectively. The same minimum is realised if $\tilde{Q}$ is present but has the same sign as $J$. However, a different phase difference is realised when $J\tilde{Q}<0$ and $|\tilde{Q}/J|\ge \frac{1}{4}$, namely $\Delta \theta=\arccos(-J/4\tilde{Q})$.

In the example of two oscillators the stationary state with equal occupancy of the nodes is always reached. However, in a more general system with many oscillators, unless the oscillatory network is highly symmetric (all oscillators have equal number of connections of the same strength with other oscillators) the systems breaks into subsystems characterised by different frequencies. To guarantee the full synchronisation of the network  we need to choose the injection rates in such a way that all oscillators have the same occupancy $\rho_{\rm th}$ \cite{kalinin2018networks}. For instance, this can be achieved by adjusting the pumping rates dynamically, depending on the occupancy of the $i-$th oscillator at time $t$:
\begin{equation}
\dot{\gamma_i}=  \epsilon(\rho_{\rm th} - \rho_i),
\label{main2}
\end{equation}
where the parameter $\epsilon$ characterizes how fast  $\gamma_i$ adjusts to changes in $\rho_i$. Aiming at the algorithmic implementation, we will focus on  tensors of the same order $k$, and leave the problems with mixed order tensors for future work. In case of polariton condensates, this means that the fourth order tensors dominate the dynamics of the second-order terms. With appropriate density adjustments, as described above,  the system of $N$ oscillators will always synchronize and achieve the stationary minimum of the Hamiltonian 
$H= -\sum_i \sum_{\langle j,k,l \rangle} \tilde{Q}_{ijkl} \cos\theta_{ijkl}$ with super-symmetric tensor of $k=4$.

To replace the minimization in  the space of continuous spins  with binary states, one could combine nonresonant pumping with resonant  
  at  twice the frequency of the condensate which introduces the terms proportional to $\psi^*$ ($\Psi_i^*$) to the right-hand side of Eq.~(\ref{gpe}) (Eq.~(\ref{ai_alt})) similarly to $k=2$ case \cite{kalinin2018networks}.  Resonant and nonresonant excitations have been previously combined in experiments on polariton condensates using chemical etching across the sample   allowing resonant excitation from the back side of the cavity \cite{ohadi2016nontrivial}.  Such combination of resonant and nonresonant   excitations would lead to the realisation of the minimum of the $k$-local Hamiltonians, so to solving Eq.~(\ref{optimisation}) with the binary spins $x_i=\cos \theta_i $, where $\theta_i$ are limited to $0$ and $\pi$. However, in contrast with $k=2$ case,  the phase projections on the binary states for $k>2$ is automatic due to the mixture of $\Psi_{i_j}$ and $\Psi_{i_k}^*$ present in the tensor form as shown below, so the presence of the additional resonant field in not necessary.   

{\it Physics-Inspired Optimization.}
 The  principle of coherence formation at a minimum of a spin Hamiltonian formulated above inspires an efficient algorithm for finding the global minimum of HOBO. For this, we extend and simplify Eq.~(\ref{ai_alt})  to capture the mechanism of relaxation to the minimum of the HOBO but without the necessity to capture  full physics of the actual system. The minimum  of HOBO for $N$ binary variables can be found by numerical integration of $2N$ equations
 \begin{eqnarray}
\frac{d\Psi_{l}}{dt}&=& \Psi_l (\gamma_l(t)  - |\Psi_l|^2) +  \sum_{\bar{\Omega}} {\bf A}_{i_1,i_2,\cdot\cdot\cdot,l,\cdot\cdot\cdot i_k}^k \Psi_{i_1} \Psi_{i_2}\cdot\cdot\cdot \Psi_{i_k}^*,\nonumber\\ 
\label{main1}
\end{eqnarray}
together with Eq.~(\ref{main2}) where $\bar{\Omega}=\Omega/ l$ and the initial values for pumping strength $\gamma_l(t=0) =-\max_{1\leq l\leq N} \sum_{\bar{\Omega}} |{\bf A}_{i_1,i_2,\cdot\cdot\cdot, l, \cdot\cdot\cdot i_k}^k|$.
At the fixed point,  the imaginary part of Eq.~(\ref{main1}) gives the set of linear equations such that the $l$-th equation involves  superposition of  $\sin(\sum_{i_j\ne \{l, {i_k}\}} \theta_{i_j}-\theta_{i_k}-\theta_l)$ that has to be equal to zero. In general, the only way  for the system to satisfy these equations is to bring all phases $\theta_l$ to take on $0$ or $\pi$. The total occupancy of the system at the fixed point is found from the real part of Eq.~(\ref{main2}) and  is equal to $N \rho_{\rm th}$, so that  $N \rho_{\rm th}=\sum_{l=1}^N \gamma_l + \sqrt{\rho_{\rm th}}^{k-2}  \sum_{{\Omega}} {\bf A}_{i_1,i_2,\cdot\cdot\cdot, i_k}^k\cos(\theta_{i_1})\cos(\theta_{i_2})\cdot\cdot\cdot\cos(\theta_{i_k}).$  If we set the process of raising the pumping from below that guarantees that $\sum_{l=1}^N \gamma_l$ is the smallest possible injected intensity, then at the fixed point the system finds the global minimum of the $k-$local Hamiltonian $H=-\sum_{\Omega} {\bf A}_{i_1,i_2,\cdot\cdot\cdot,  i_k}^k\cos(\theta_{i_1})\cos(\theta_{i_2})\cdot\cdot\cdot\cos(\theta_{i_k}),$  and, therefore, solves Eq.~(\ref{optimisation}). We will refer to the Eqs.~(\ref{main1}-\ref{main2}) as  the Tensor Gain-Dissipative (TGD) method.  

\begin{figure}[!h]
		\includegraphics[width=8.2cm]{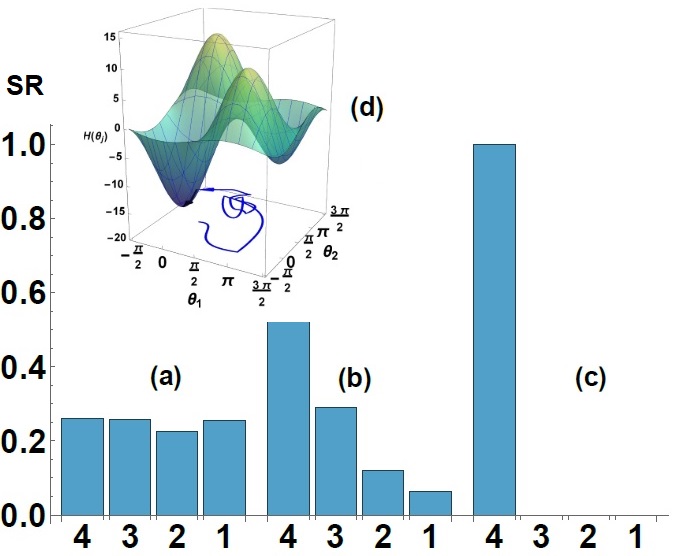}
		\caption{Success rates for achieving local and global minima of $H_{\rm test}$  using numerical simulations of Eqs.~(\ref{main1}-\ref{main2}) using fully deterministic integration without noise (a); with white noise added to the right hand side of Eq.~(\ref{main1}) (b); using complex coupling control as described in the main text (c). The indexes marked on the horizontal axis label local minima as $1,2,3$ that correspond to $H_1, H_2$ and $H_3$ respectively. The global minimum ($H_4$) is labeled $4$. The insert (d) shows
the two-dimensional projection ($\theta_3=\theta_4=0$) of the energy landscape for $H_{\rm test}$ with $x_i=\cos \theta_i$. The  time evolution of the scaled  total injected intensity $(\sum \gamma_i-N\rho_{\rm th})/3$ is   shown by the blue trajectory. }
		\label{statistics}
\end{figure}

To illustrate the behavior of the system we first consider a toy problem: the following 3-local  Hamiltonian
%
$H_{\rm test}(\textbf{x}) = - 8 x_{1} x_{2} x_{3} - 4 x_{1} x_{2} x_{4} - 2 x_{2} x_{3} x_{4} - x_{1} x_{3} x_{4}, $
with variables $x_i\in\{\pm 1\}$, while Eq.~(\ref{main1}) becomes $\dot{\Psi_l}=\Psi_l(\gamma_l-|\Psi_l|^2) + \sum_{\langle j,k\rangle}K_{ljk}\Psi_j\Psi_{k}^*,$ and $K$ is a tensor with nonzero entries $1,2,4,8.$ 
The Hamiltonian $H_{\rm test}$ has $2^4$ stationary points, among which there are three local minima: $H_1=-9,H_2=-11, H_3=-13$  and the global minimum $H_4=-15$, that all can be accessed during the time evolution of the system. To understand the basins of attraction for these stationary points we numerically integrate Eqs.~(\ref{main1}-\ref{main2}) starting with  initial conditions $\Psi_i(t=0)=\frac{1}{100}\exp[{\rm i} \theta_{0i}]$ where the phases $\theta_{0i}\in[0,2\pi)$ are uniformly distributed  \cite{numerics1}. Figure~\ref{statistics}(a) depicts the statistics of distribution of the stationary points reached and indicates that the basins of local minima combined  are larger than that of   the global minimum. To facilitate the search for the global minimum the algorithm needs to allow for a possibility to explore the hyperspace until the lowest lying energy state is found. This can be achieved by adding a Langevin noise, which represents intrinsic vacuum fluctuations and classical noise, that shifts the trajectory from its deterministic path while allowing it to stay below any local minima. 
Figure \ref{statistics}(b)  depicts the statistics of reaching local and global minima found by numerical integration of Eqs.~(\ref{main1}-\ref{main2}) using the same initial conditions as in Fig.~\ref{statistics}(a) but with the white noise added.  Decreasing the $\epsilon$ parameter allows to improve the possibility of reaching the global minima. Figre \ref{statistics}(d)  shows one  such trajectory as it approaches the global minimum of $H_{\rm test}$ from below. 

With the growth in the number of variables and concomitant growth of the system hyperspace  local noisy perturbation of the trajectory may not be sufficient to reach the global minimum basin of attraction.  Motivated by recent studies of  heteroclinic networks \cite{neves2012computation} we introduced heteroclinic orbits into our model by engineering time-dependent complex couplings into the network Eqs.~(\ref{main1}-\ref{main2}). Complex couplings naturally appear in polariton model if the energy shift due to a noncondensed reservoir $R({\bf r},t)$ is present in the system \cite{kalinin2019polaritonic}.  The imaginary part of the coupling may destabilise the stable fixed point so that the system trajectory quickly leaves  its neighborhood  along the fastest direction. Including this switching dynamics into the system facilitates the search for the true global minimum by allowing fuller exploration of the phase space.

{\it Complex coupling switching}. 
To implement the complex coupling switching method (TGD+CC) on  $H_{\rm test}$ we turn two of the real coupling coefficient into the complex ones with a significant imaginary part as soon as the system reaches a steady state. The system trajectory leaves the basin of attraction of that state and travels to a different part of the system hypercube $[0,2 \pi]^N$. When the imaginary part of the coupling is turned off another steady state will be found.   
By varying the coupling elements to be switched, the time duration of the switching, and the  amplitude of the imaginary part  while keeping the injected intensity low we allow the system to efficiently search for a low energy minimum. In our test example, implementing the switching of a coupling coefficients $K_{123}$ and $K_{124}$ according to $K_{123}(t)=8(1+4i), K_{124}(t)=4(1-10i), t\in [t_1,t_1+160] \cup [t_2,t_2+160] \cup [t_3,t_3+280]$ and keeping $K_{123}(t)=8, K_{124}(t)=4$ otherwise allows {\it every} trajectory irrespective of its initial state to reach the global minimum, see Fig.~\ref{statistics}(c) . Here $t_1, t_2, t_3$ are times at which the system settles to a steady state after switching off the imaginary part of the couplings \cite{numerics2}. 
\begin{figure}[!h]
		\includegraphics[width=8.6cm]{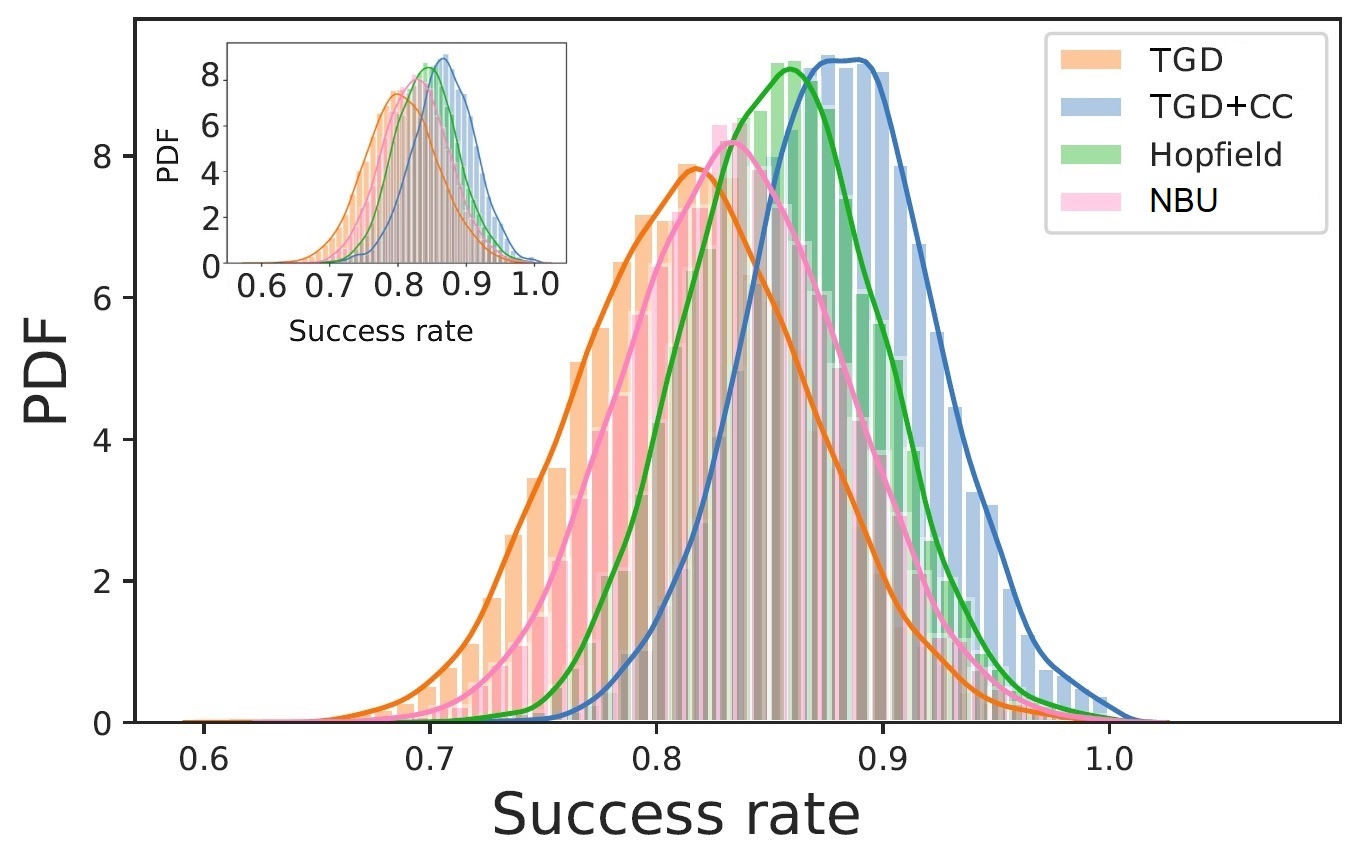}
		\caption{Success probability density distribution  for $500$ realisations on the set of $20$ dense (main) and sparse (inset) $3^d$ rank tensors $K_{ijk}$  with $10^6$ elements for four different methods discussed in the main text. To generate sparse tensors 9/10th of all elements in dense tensors were set to zero.    }
		\label{histograms}
\end{figure}

{\it Complex coupling switching for large N}.
We tested the complex couplings switching approach using the large scale simulations   on $20$ dense and $20$ sparse random tensor sets $K_{ijk}$ of $3^{d}$ rank  with $10^{6}$ elements over $500$ different realisations.  To implement TGD+CC method on large $N,$ as soon as the system reaches the steady state we randomly choose $N/50$ of the coupling strengths $K_{ijk}$  and modify them by adding  $3iK_{ijk}$. As the system trajectory leaves the basin of attraction of the previous fixed point, we let return to original couplings and allow the system to relax to a new steady state. Keeping the total injected rate $\sum \gamma_l$ small forces the trajectories to explore the low energy states of the Hamiltonian until the true global minimum is found  \cite{numerics_bign_withcc}.

We compare the behaviour of the TGD and  TGD+CC with  two popular network-based methods and show that TGD+CC  outperforms all of these methods. 
The first  method represents the network of analogue  bistable  units  (NBU) in the presence of a double-well potential derivative that forces the network elements $x_l$ to take on $\pm 1$ while solving  Eq.~(\ref{optimisation}):
\begin{eqnarray}
\frac{dx_{l}}{dt}&=& -h x_l |x_l|^{k-1} (x^{2}_l  - 1) +  \sum_{\bar{\Omega}} {\bf A}_{i_1,i_2,\cdot\cdot\cdot,l, \cdot\cdot\cdot, i_k}^k x_{i_1} x_{i_2}\cdot\cdot\cdot x_{i_k}\nonumber,\\&\label{gradient1}
\end{eqnarray}
where  $x_{l}(t=0)$ are randomly distributed real numbers, and $h$ is a control parameter \cite{numerics_gradient}. In comparison with the usual $k=2$ case \cite{leleu19}, we balanced the degrees of polynomial between two term on the right-hand side of Eq.~(\ref{gradient1}) by introducing $|x_l|^{k-1}$ factor. 
Another efficient solver of Eq.~(\ref{optimisation}) is given by a higher order Hopfield neural networks  \cite{joya2002hopfield}:
\begin{equation}
\frac{dx_{l}}{dt}= -x_l +  \sum_{\bar{\Omega}} {\bf A}_{i_1,\cdot\cdot\cdot,l,\cdot\cdot\cdot, i_k}^k \tanh\biggl(\frac{x_{i_1}}{\beta}\biggr) \cdot\cdot\cdot \tanh\biggl(\frac{x_{i_k}}{\beta}\biggr),
\label{hopfield1}
\end{equation}
where $x_{l}$ are real continuous variables and $\beta$ is the scaling parameter \cite{numerics_hopfield}. 

Figure \ref{histograms} shows the results of large-scale numerical simulations using different methods. TGD+CC consistently has a better success probability of finding the global minimum.

{\it Conclusions.} 
In this Letter we showed that lattices of nonequilibrium condensates when pumped above the threshold  may realise $k$-local Hamiltonians  with nontrivial spin structures. We  formulated system-inspired method of computing the optimal solution of a large range of HOBO problems. Finally, we introduced the concept of computation
via the mechanism of  complex coupling switching. Its combination with  the tensor variation of the gain-dissipative algorithm  leads to an efficient way of finding the low energy states of the Hamiltonian due to: (i) individual node gain control that allows to explore  low energy states of the Hamiltonian and  guarantees the achievement of the minimum, (ii) evolution in real number space that allows to tunnel through functional barriers in discrete variables, (iii)  Kuramoto networks graduate decent close to threshold, and finally (iv) complex coupling switching  that allows the trajectory to escape local minima.  This approach offers a highly flexible new kind of computation that is inspired by and compatible with  many physical network realisations. We envision it becoming a part of a hybrid platform where the states of the network are fed to the physical device for the optimal performance.

The authors acknowledge support from Huawei.

	
	\bibliography{Bibliography}{}

\begin{thebibliography}{10}
\bibitem{numerics1}
Eqs.~(\ref{main1}-\ref{main2}) were Euler integrated with $\epsilon = 0.001M$,
  $\gamma_{l}(t=0)=-0.2M$, $\rho_{\rm th}=1,$ with $D=\max(1-\rho/\rho_{\rm
  th}, 0)$ , $dt=0.01$, $T_{\rm stop}=300$, $M=\max_{1\leq l\leq N}
  \sum_{\bar{\Omega}} |A_{l,i_1,i_2,\cdot\cdot\cdot, i_k}^k|$. Tensor elements
  were scaled by $0.05$ to slow the dynamics down.
\bibitem{numerics2}
The switching times are defined by analysing the transitions between the steady
  states labeled by $1,2,3,4$ as described in the main text. Turning on the
  complex part of $K_{123}(t)=8(1+4i)$ and $K_{124}(t)=4(1-10i)$ for $t=T$
  after the steady states are reached leads to the following permutation of the
  states: a)$(1,2,3,4)\rightarrow (2,3,3,4)$ if $T=80;$ b)$(1,2,3,4)\rightarrow
  (1,1,1,2)$ if $T=160;$ c)$(1,2,3,4)\rightarrow (4,3,3,4)$ if $T=280.$ Clearly
  the switching protocol b)b)c) brings all trajectories to the global minimum.
\bibitem{numerics_bign_withcc}
Eqs.~(\ref{main1}-\ref{main2}) were Euler integrated with $\epsilon = 0.01M$,
  $\gamma_{l}(t=0)=-0.2M$, $\rho_{\rm th}=0.1M,$ with $D=100
  \max(1-\rho/\rho_{\rm th}, 0)$, $dt=0.001$, $M=\max_{1\leq l\leq N}
  \sum_{\bar{\Omega}} |A_{l,i_1,i_2,\cdot\cdot\cdot, i_k}^k|$. The elements of
  dense (sparse) tensors were multiplied by $5/6N$ ($15/6N$).
\bibitem{numerics_gradient}
Eq.~(\ref{gradient1}) was Euler integrated with $dt=0.001$ with the saem number
  of iterations as in other methods. The initial conditions $x_l(t=0)$ were
  uniformly randomly distributed in $[-0.5,0.5]$, and $h_{j+1}=1.2h_{j}$ is
  updated each time $\sum_{l}|\frac{dx_{l}}{dt}|<0.001$ is satisfied.
\end{thebibliography}
	\bibliographystyle{ieeetr}

\end{document}